\documentclass[pra,aps,showpacs,a4paper]{revtex4}
\usepackage[matrix,frame,arrow]{xypic}
\input{Qcircuit}
\usepackage{graphics}
\usepackage{times,mathptm}
\usepackage{epsfig}
\usepackage{mathrsfs}
\usepackage{amsmath,amssymb}
\usepackage[colorlinks]{hyperref}
\usepackage{subfigure}

\newcommand{\beq}{\begin{equation}}
\newcommand{\eneq}{\end{equation}}
\newcommand{\beqy}{\begin{eqnarray}}
\newcommand{\eneqy}{\end{eqnarray}}

\begin{document}

\title{Fault tolerant Quantum Information Processing with Holographic control}
\author {Gerardo~A. Paz-Silva}
\altaffiliation{E-mail: {\tt gerardo.paz-silva@mq.edu.au
}}
\author {Gavin K. Brennen}
\author {Jason Twamley}

\affiliation{Centre for Quantum Computer Technology, Macquarie University, Sydney, NSW 2109, Australia}

\date{\today}

\begin{abstract}
We present a fault-tolerant semi-global control strategy for universal quantum computers. We show that N-dimensional array of qubits where only (N-1)-dimensional addressing resolution is available is compatible with fault-tolerant universal quantum computation. What is more, we show that measurements and individual control of qubits are required only at the boundaries of the fault-tolerant computer, i.e. holographic fault-tolerant quantum computation. Our model alleviates the heavy physical conditions on current qubit candidates imposed by addressability requirements and represents an option to improve their scalability.
\end{abstract}


\maketitle

\section{Introduction}

In recent years much effort has been devoted to the realization of what was first envisioned by Shor and collaborators~\cite{Shor}: a quantum computer. To achieve this one basically requires: a set of qubits, and the ability to prepare, manipulate (through gates) and measure quantum information stored in those qubits~\cite{DiVincenzo}. In the face of errors one must also execute fault-tolerant quantum error correction (QEC) to enable large scale quantum information processing~\cite{QEC}. For fault-tolerant universal quantum computing (FTUQC)~\cite{threshold} one typically needs some high degree of parallelism or simultaneity. In particular, traditional schemes require  (i) complete addressability, (ii) sufficiently low error rates for operations, and (iii) simultaneity of such operations. Such tough requirements imply that the experimental execution of fault-tolerant schemes will be technically very challenging. For example, ion-traps have fairly low error rates ($\sim 10^{-3}$)) and good addressability but scaling ion traps to large number of ions where many gates can be executed simultaneously is highly challenging~\cite{IonTap}. On the other hand, neutral atoms trapped in optical lattices posses  large numbers of qubits and also simultaneous control but individual addressing is very challenging~\cite{OLatt}. In general there is a trade-off between the capability to perform individual addressing and large-scale simultaneous addressing. This leads us to ask the questions: which of the requirements (i-iii) can be relaxed and still achieve FTUQC with a reasonable threshold? Relaxing the requirements may pave the way for implementations to scale up the number of qubits and achieve fault-tolerant quantum computation. 

In a previous paper~\cite{noisymeas} we approached the problem trying to reduce the requirements on criteria (ii) the accuracy of operations. One can reduce, in principle, the addressing requirements using global control techniques. In global control, gates don't have to be individually addressed and one can achieve universal quantum computing using global operations. The preparation of specific quantum states can be replaced with a resetting process plus a gate, moreover the resetting process can be executed in parallel, greatly reducing the addressing requirements. Only measurements seem to require addressing. This apparent requirement poses heavy experimental constraints when one demands fault-tolerance: error correction in every logical qubit simultaneously, at every computational time-step, which means that measurement-aided syndrome extraction would require one to simultaneously and distinguishably measure the state of many qubits. Distinguishable measurements with good accuracy may be possible in small qubit registers but will be very challenging to scale up~\cite{NatureHardMeasurements}. Consequently, to avoid these and other issues, in Ref.\cite{noisymeas} we introduced circuits to remove measurements from quantum error correction protocols and greatly reduce the role of measurements in fault-tolerant universal quantum computing. Our results showed two things: (I) gates must be quite accurate while preparation and measurements can have error rates as high as 33\%; (II) because measurements are removed, simultaneous and paralell FT QEC can be executed over many logical qubits. 

In this paper we explore deeper into the consequences of (II) and propose a novel concept with the potential to greatly alleviate not only the measurement-related issues mentioned above but also reduce some of the addressability constraints which typically burden fault-tolerant computer designs. We show for the first time that fault-tolerant holographic, i.e. addressing the boundaries only, universal quantum computing is possible. We will argue that such holographic design not only represents a way of reducing the number of individually addressable qubits but also inherits the results we obtained in Ref.\cite{noisymeas} regarding the tolerable high error rates for fault-tolerance,

The paper is organized as follows. In section \eqref{HoloFTUQC} we discuss the addressing required in typical FTUQC designs, and describe the schematic idea of a holographic, i.e. semi-global with boundary addressing only, control scheme which reduces the number of controls and amounts of addressing required while maintaining fault-tolerance. In section \eqref{HFTUQC} we formalize the scheme and explicitly describe the necessary tools. In particular we describe fault-tolerant routines which make minimal use of measurements; these measurements can be done exclusively on the boundary i.e. holographically (or offline). We then show how our model can be mapped to a lower dimensionality array but now requiring additional next-to-nearest neighbor gates. We discuss (Section \eqref{secthresh}) the error threshold for this design and finally in section \eqref{controlsec} we proceed to make an analysis of the resources required and show that the number of controls in our strategy has only a weak dependence with the number of computational qubits, as opposed to traditional fully addressable architectures. 

\section{A semi-globally controlled quantum computer}
\label{HoloFTUQC}
We initially consider a 3D array of qubits with nearest neighbor interactions although we urge the reader to keep in mind that our real goal is to show that a N-dimensional, $N=2,3$, array can execute fault-tolerant quantum when only (N-1)-dimensional individual addressing is available. We will first develop the 3D model and then, in section \eqref{sec2D}, argue that the model can work in a 2D array with 1D individual addressing. A 3D array of qubits makes efficient use of space, however in a 3D spatial array it will typically be hard to measure/manipulate individual qubits in the bulk of the array, i.e. 3D addressing resolution. Instead we assume a 2D addressing resolution, i.e. the ability to address lines of qubits in the array. We label indexed array of 3D locations by coordinates $(x,y,z)$, where each coordinate $s \in [1,N_s]$, for $s =\{x,y,z\}$ and label the addressable lines in the 3D array by $(x,y)$. The action of a {\it single-qubit} gate addressing the line $(x,y)$ is given by $U_{(x,y)} = \prod_{z} U_{(x,y,z)}$, while {\it two-qubit} gates between addresses $(x,y)$ and $(x',y')$ is given by $V_{((x,y),(x',y'))} = \prod_{z} V_{(x,y,z),(x',y',z)}$ , with the obvious generalization for multi-qubit gates. Approaching measurements this way is not practical as it does not allow one to discriminate individual qubit measurement results along $z$. We shall assume that all measurements are executed at the boundaries. In fact we allow the possibility of executing any operation on the $z$-boundaries i.e. $O_{(x,y,1)}$ and $O_{(x,y,N_z)}$ for any $(x,y)$. The addressing limits described above impose a constraint on the type of gates we can execute in the $z$ direction: for example, we cannot directly execute a gate of the form $V_{(x,y,z),(x,y,z')}$. We note that we allow long-range interactions within every $z$-plane but that one can restrict to nearest neighbor gates with a slight reduction of the fault tolerant threshold due to the introduction of intermediate SWAP gates. We require the ability to execute nearest neighbor CZ gates in the $z$ direction along $(x,y)$ columns. We will show that this limited addressability, where we can only address columns of qubits in the 3D array, yield universal quantum computing and, more importantly, fault-tolerance. 

\begin{figure}[htbp]
	\centering
		\includegraphics[width=0.6 \columnwidth] {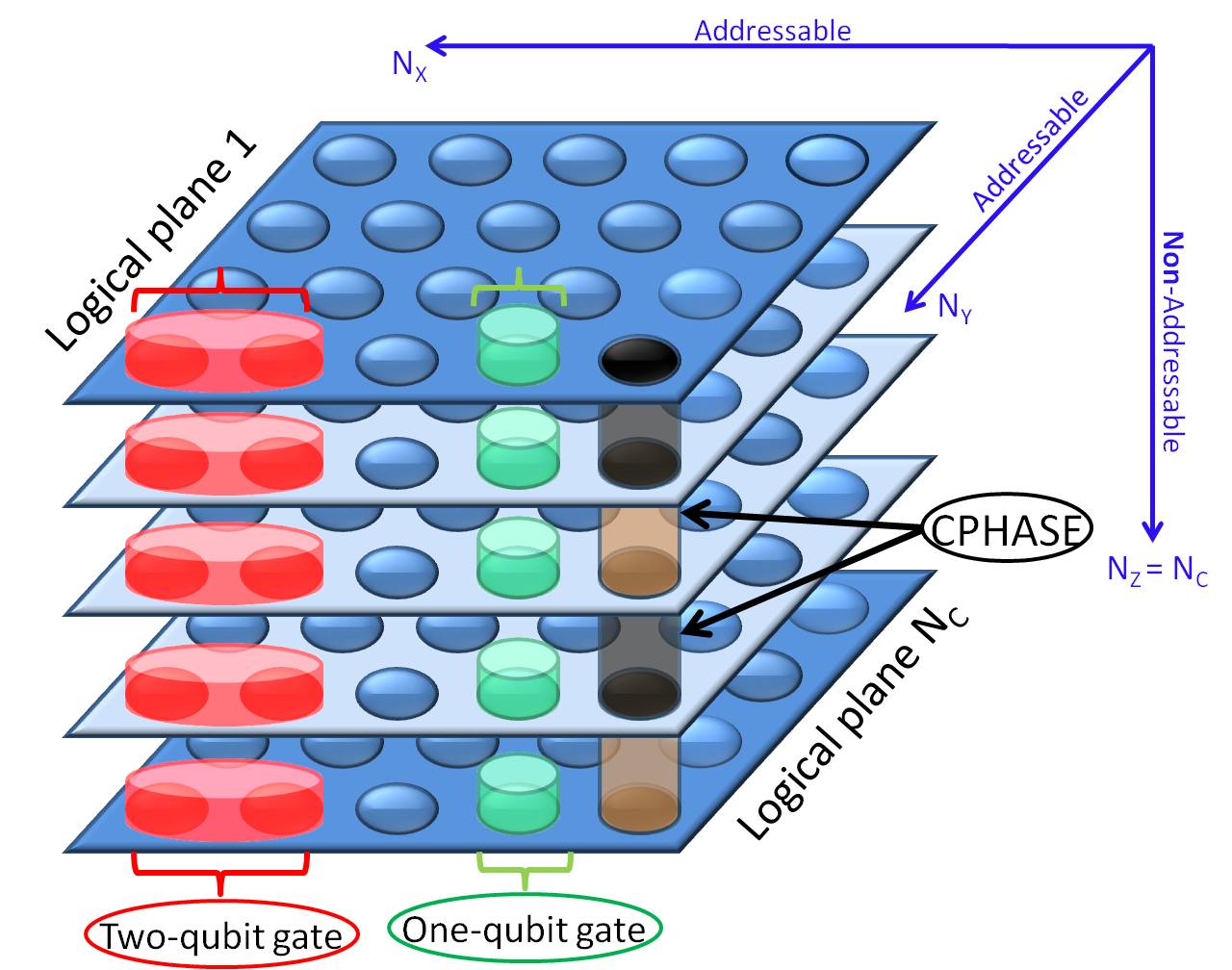}
	\caption{Schematics of the addressability requirements and qubit distribution of our semiglobal architecture. Vertical global pulses are capable of executing the {\it single-qubit} (green) and {\it two-qubit} gates (red) described in the text. Computation is achieved through the vertical nearest neighbor independent $\tilde T$ pulses {which can be decomposed into the two subroutines (black and brown $\tilde T$ pulses), in virtue of the ABAB addressability, so as to ensure fault-tolerance}. The end-planes (darker blue) are fully addressable independently of the bulk of the computer, and must have extra space to accommodate a $\ket{H_L}$ state encoded at the highest level of concatenation in order to execute the non-Clifford encoded gates. All planes plains contain enough qubits to hold an encoded qubit, the ancillas required for its unitary UQEC. We also require that every plane has physical qubits that can be reset (simultaneously in all planes) to use a resource for algorithmic cooling (simultaneous in all planes), note that the reseting operation assumes no single plane-addressability.}
	\label{semiglobal}
\end{figure}

\subsection{Global control} 
Lets consider the global control model introduced in ~\cite{FitzsimonsTwamley} (which is closely related to ~\cite{Raussendorf}). In~\cite{FitzsimonsTwamley} the authors consider a one dimensional array of qubits, e.g. consider the single column $x=1=y$ in our 3D array show that with the global operators
\beq
\label{transp}
\widetilde{CZ} = \prod_{z=1}^{N_z-1} CZ_{((1,1,z),(1,1,z+1)},\,\,\,\widetilde{H}  = \prod_{z=1}^{N_z} H_{(1,1,z)},
\eneq 
and non-Clifford gates executed at the edges of the 1D array ($z=1, N_z$), universal quantum computation is possible within this 1D array. Their basic idea uses the global gate $\widetilde T = \widetilde{CZ} \cdot \tilde{H}$ to propagate information in a controlled way within the 1D array. $\widetilde T^{N_z+1} $ is equivalent to executing a spatial reflection of the information stored in the 1-D sub-array along the z-direction: $\widetilde T^{N_z+1}:\rho_{(1,1,z)} \rightarrow \rho_{(1,1,N+1-z)}$, where $\rho$ is the density matrix of the array. By executing Clifford and non-Clifford gates at the boundaries of the 1D array and between the $\widetilde T$-pulses it was shown that universal quantum computing is possible. We refer the reader to Ref.~\cite{FitzsimonsTwamley} for details.

The problem with global control is its apparent incompatibility with fault-tolerance. The use of global pulses, usually in the form of nearest neighbor gates leads to two serious obstacles towards implementing fault tolerant QEC: (i) traditional global control techniques may give rise to correlated errors in a codeword, which although not necessarily lethal for fault-tolerance, are known to reduce the error correcting capabilities of a code~\cite{GBanacloche}, and, perhaps the most relevant, (ii) the uncontrolled propagation of errors to multiple locations e.g. due to the interaction of a faulty qubit with others.  

Although some evidence for the existence of a fault-tolerant threshold using global addressing exists~\cite{FTGCthresh}, no such threshold has been calculated to date. In any case, as global control uses many global operations to implement a single logical gate, globally simulating traditional error correction circuits can only lead to a worse threshold. 

Rather than propose a fully globally addressed architecture we here propose to use a hybrid strategy where, while the addressing requirements are reduced, fault-tolerance is still possible. The central concept of our hybrid design is to $trap$ any possible correlated error so that it won't affect more than one qubit in every  logical qubit (or more than one depending of the QEC code we are using, for simplicity we will assume in this paper distance 3 codes). More importantly, we restrict the direction in which errors can propagate, ensuring that correlated errors propagate within separate codewords. {To arrange for this we consider a 3D cubical array of physical qubits, where each $xy$-plane contains a logical qubit encoded in a CSS code e.g. Bacon Shor code, Steane code, etc. The $encoded$ $T$ gate is then a set of vertical, nearest neighbor gates, in the $z$-direction which, given the choice of code, are transversal, i.e. bitwise,
\beq 
T = \prod_{(x,y,z)} CZ_{(x,y,z),(x,y,z+1)} H_{(x,y,z)} = \prod_{(x,y)} \tilde {T}_{(x,y)},
\eneq
where $\tilde {T}_{(x,y)} = \prod_{z} CZ_{(x,y,z),(x,y,z+1)} H_{(x,y,z)}$. We note that a single faulty $physical$ $\tilde{T}_{(x,y)}$, i.e. any combination of errors in the gates composing $\tilde{T}_{(x,y)}$, can generate a correlated error, but such error will only affect $one$ qubit in every plane, i.e. in every logical/encoded qubit. Furthermore, to avoid simultaneous $CZ$ gates targeting or controlling more than one logical plane we execute $T$ in two steps to get a fault-tolerant version, $\mathcal{T}$, 
\beq \label{ftT} \mathcal{T} = \widetilde{CZ}^{o-e}_{(x,y)} \times \widetilde{CZ}^{e-o}_{(x,y)} \times \prod_{(x,y,z)} H_{(x,y,z)},\eneq
where $\widetilde{CZ}^{o-e}_{(x,y)}= \prod_{(x,y,z \in [1, \lfloor N_z/2 \rfloor])} CZ_{(x,y,2n-1),(x,y,2n)}$ and $\widetilde{CZ}^{e-o}_{(x,y)}= \prod_{(x,y,z \in [1, \lfloor N_z/2 \rfloor])} CZ_{(x,y,2n),(x,y,2n+1)}$. This would require an ABAB plane addressability where we are able to execute AB and BA nearest (plane) neighbor operations. Moreover if we use the Bacon Shor code, the CZ is not completely transversal but requires an extra $\pi/2$ physical rotation of one of alternate (x-y) planes in the 3D array. To avoid this, particularly for the BS code we benefit from the ABAB plane addressability: (i) even planes will be encoded in the BS code while odd planes are encoded in a $\pi/2$ rotated BS code and (ii) even and odd planes will have independent error corrections according to the standard or rotated BS encoding. Other CSS codes, such as the Steane code, won't have this issue as the CZ gates are transversal in this code but will still require the ABAB addressability for the execution of the fault-tolerant $\mathcal{T}$.}

\subsection{Quantum error correction without measurements (UQEC)}
\begin{figure}[h]
\subfigure[]{
\includegraphics[width=0.4 \columnwidth]{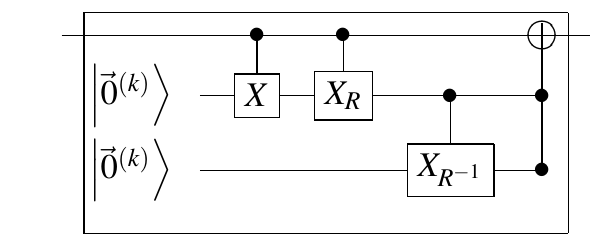}       
\label{M-gate}
}
\subfigure[]{
\includegraphics[width=0.4 \columnwidth ]{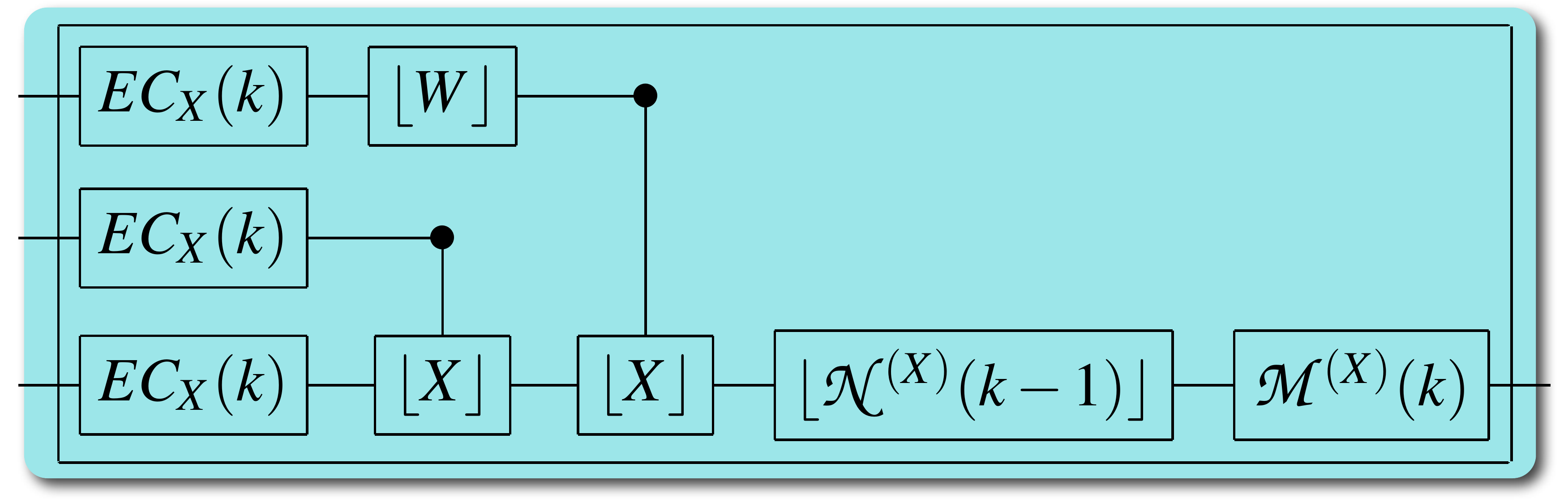}
\label{VN}}
\subfigure[]{
\includegraphics[width=0.8\columnwidth]{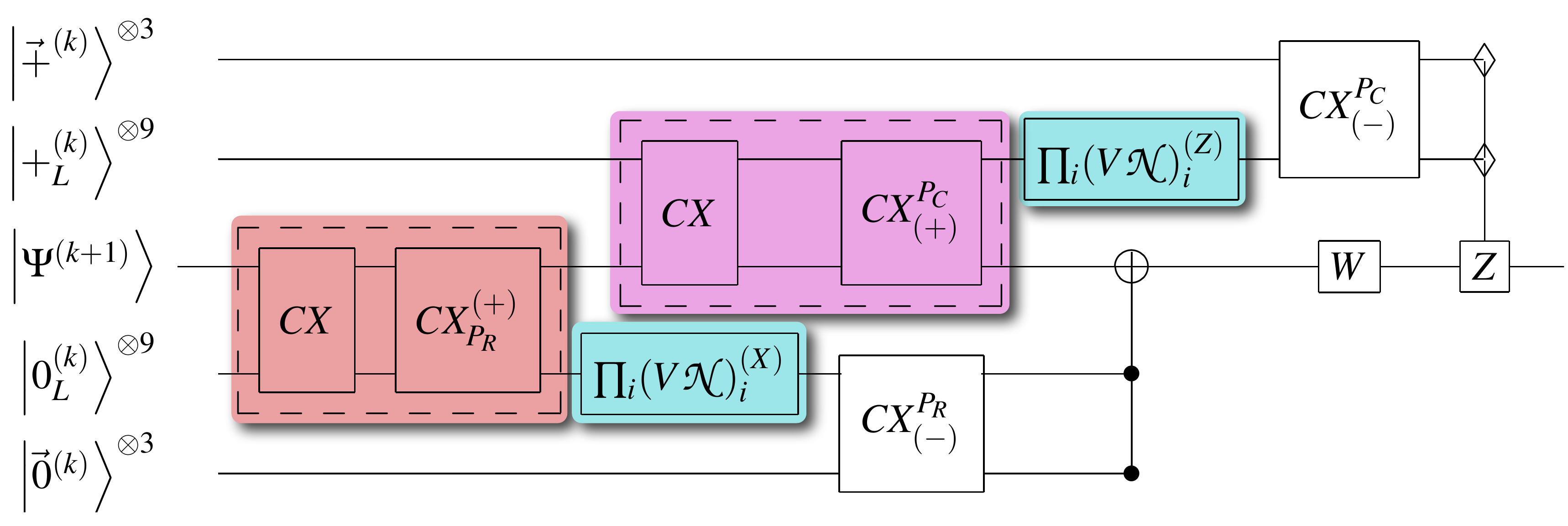}
\label{BS}}

\caption{ Measurement-free QEC routines for the QR and BS code. The inputs are $\ket{\vec 0^{(k)}}= \ket{000}^{\otimes 3^{k-1}}$ and $\ket{\vec +^{(k)}}= \ket{+++}^{\otimes 3^{k-1}}$.  
\textit{(a) The $\mathcal{M}$ gate.} An X-encoded majority voting gadget of level-$(k+1)$ of concatenation. Here all CNOTs are bitwise, i.e. each CNOT depicted corresponds to three $CNOT(k)$, and subscript $R$ corresponds to a cyclic $k$-encoded rotation of the targets of the corresponding gate. In the QR code the TOFFOLI gate depicted is bitwise. The $\mathcal{M}$ gate can also  
be designed for a Z-encoded quantum majority voting, with $\ket{\vec{+}^{(k)}}$  
ancillas and the obvious Hadamard conjugation of gates. When the need  
to distinguish them arises we shall denote $X$ and $Z$ encoded  
majority votings $\mathcal{M}^{(X)}$ and $\mathcal{M}^{(Z)}$  
respectively.
\textit{(b)} A subroutine acting on ancilla for processing error syndrome information extracted from the data.  The circuit shows one row, $(V\mathcal{N})_i (k)$, of the fully contracted exRec $V\mathcal{N}(k)$ representing a collection of $k$-level protected gates acting on row $i$ of ancilla which take part in an $EC(k+1)$ step. Note that in this circuit the output top lines are discarded so no EC gadget must protect them. With this, the exRec corresponding to $\mathcal{N}$ at degree of concatenation $k$ is $\mathcal{N}(k) = EC(k) \times \prod_{i \in rows} (V\mathcal{N})_i (k-1) \times EC(k)$. In our circuits $\lfloor G(k) \rfloor$ denotes the implementation of gate $G$, in terms of level-$(k-1)$ gates, without the prepending or appending $EC(k)$.
\textit{(c) Full error correction (EC) gadget for the BS code.} The orange and pink boxes represent the syndrome extraction stage. Here, a  
TOFFOLI with $\diamond$ controls is a $Z-TOFFOLI$; $CX =  
\prod^{3}_{i,j=1} CNOT^{(k)}_{(c,i,j),(t,i,j)}$ is a set of transversal  
CNOTs, $CX^{(\pm)}_{P_R} = \prod^{3}_{i,j=1} CNOT^{(k)}_{(c,i,j),(t,i,j \pm 1)}$
and $CX_{(\pm)}^{P_C} = \prod^{3}_{i,j=1} CNOT^{(k)}_{(c,i,j\pm 1),(t,i,j)}$. The control of the gates in boxes is always the top input of the gate. The $W$ gate is a wait (identity) gate, and the last gate on the upper and lower half is a transversal bTOFFOLI.}
\end{figure}

We now consider how to execute quantum error correction routines given the limited in-plane addressability. Our restriction to columnar semi-global operations forbid the the execution of parallel individual measurements along the $z$-direction. This would indicate that parallel, measurement-aided, error correction and thus fault-tolerance is not possible. The straightforward solution is to remove measurements from quantum error correcting routines, and use suitably designed circuits instead. In Ref.~\cite{noisymeas} we introduced such unitary error correcting routines for the BS code (Fig.\eqref{BS}) based on the so-called $\mathcal{M}$-gate: a majority voting gadget (Fig.\eqref{M-gate}). Although the specific design we use here for the M-gate is tailored for the BS code we stress the measurement free scheme below applies independently of this choice of QEC code (See supplementary material in Ref. \cite{noisymeas}). Armed with the appropriate unitary quantum error correction (UQEC) gadgets and certain transversality properties (which we outline below) the semiglobal scheme can be applied to a variety of CSS~\cite{CSS} codes. 

The error correction scheme used here is based on two routines: an error correction routine for the QR code, the $\mathcal{M}$, and a mapping between CSS and QR codes,  using a gadget we dub the $\mathcal{N}$-gate. The Bacon-Shor (BS) code is a composition of, X- and Z- base, quantum repetition (QR) codes defined by the following stabilizer set on a two dimensional array of 3x3=9 qubits: 
\beq
S = \left\{ \begin{array}{ccc} X&X&I\\X&X&I\\X&X&I  
\end{array},\;\; \begin{array}{ccc} I&X&X\\I&X&X\\I&X&X \end{array}, \;\; 
\begin{array}{ccc} Z&Z&Z\\Z&Z&Z\\I&I&I \end{array},\;\;\begin{array}{ccc}  
I&I&I\\Z&Z&Z\\Z&Z&Z \end{array} \right\}.
\eneq
For this code the logical Pauli operators are given by $X_L =  
\prod_{i=1}^{3} X_{i,1}; Z_L = \prod_{i=1}^{3} Z_{1,i}$ modulo  
stabilizer operations, i.e. $X_L$ ($Z_L$) acts on a column (row) of the array. 
This code is a subsystem code and is invariant under pairs of $X(Z)$ operators along any given row(column) because they act only on gauge degrees of freedom. Given the subsystem structure of the code one is able to correct acting on only one row (for X-errors) and on only one column (for Z-errors). An encoded $\mathcal{M}$-gate provides a way of executing BS error correction. Here is where the $\mathcal{N}$-gate comes into play: as an interface between the BS and the QR codes. Let us explain how the UQEC protocol operates. For illustration lets assume we are executing the X-correction stage i.e. lower part of Fig.\eqref{BS}. We can extract the syndrome information using BS encoded gates at any level of concatenation; then we use the $\mathcal{N}$-gate to take a BS encoded syndrome to a QR encoding. This is possible because the syndrome used in a particular error correction stage only needs protection against one type of error, e.g. during the X-stage, the syndrome must be protected against $X$ errors, and not $Z$ errors. After this step we have three syndrome strings, one for each column of the encoded $3\times 3$ array, $(s1,s2,s3)$. If we tried to correct every column we would destroy the gauge-freedom available for the BS code, so in order to maintain gauge freedom we use these strings to vote into a fourth $s_4 = s_1 \oplus s_2 \oplus s_3$, which will control the final correction. This step is achieved through the $V\mathcal{N}$ routine described in Fig.\eqref{VN}. Note that the $s_4$ string is encoded in the QR code, so it may seem that we won't be able to execute the correction on a BS encoded state. However the interplay between CSS and QR codes is very interesting: a CNOT gate and more importantly a TOFFOLI gate can be executed using QR encoded controls and can target a subset of the CSS encoded state, e.g. targeting one column of the BS encoded state. To actually execute the $s_4$ correction we just copy it with a cyclic rotation in order to execute the correction through a bitwise TOFFOLI gate. To see how the voting respects the gauge freedom consider the following scenarios: a single error in e.g. column one, leads to $s_4 = s_1$, which would correctly execute the correction by virtue of the gauge freedom; on the other hand a gauge-like operation, two $X$-errors in the same row, leads to $s_4 = s \oplus s = \vec{0}$, which correctly implies an identity correction operation. An analogous analysis holds for the $Z$-error correction. This completes our schematic description of the BS UQEC gadget.

To complete a UQEC scheme capable of achieving fault-tolerance we must provide fresh ancillas in every $(x-y)$ plane. Fresh ancillas can be generated via a semi-global reset, which in contrast with measurement, does not need to output a result and thus imposes no addressability constraints. A noisy version of this operation with fails with probability $\tilde p_{(p)}$ can be modeled by 
\beq 
\rho \rightarrow (1-\tilde p_{(p)}) \ket{0}\bra{0} + \tilde p_{(p)} \eta,
\eneq 
where $\rho$ is an arbitrary density (of a qubit in each x-y plane) matrix ideally mapped to $\ket{0}\bra{0}$ state, or to an arbitrary density matrix $\eta$ with probability $\tilde p_{(p)}$. If we require an even lower error rate, then we can  further use an algorithmic cooling protocol \cite{noisymeas} in parallel to distill colder $\ket{0}$ states and effectively reduce it, $\tilde p_{(p)} \rightarrow p_{(p)} < \tilde p_{(p)}$. With this fresh ancillae we can also unitarily prepare the level-k BS encoded states, $\ket{0^{(k)}_L}$ and $\ket{+^{(k)}_L}$, the level-k QR encoded states, $\ket{\vec{0}^{(k)}}$ and $\ket{\vec{+}^{(k)}}$, needed for error correction at every level of concatenation, using the following routines.
\begin{itemize}
\item Preparation of $\ket{\vec{0}^{(k)}}$ and $\ket{\vec{+}^{(k)}}$: 
Using three copies of $\ket{0}$ we execute and $\mathcal{M}^{(X)}$ gate to ensure one has a $\ket{\vec{0}}$. In reality the $\mathcal{M}$ gate is taking the role of the error correction gadget. This routine can be concatenated any number of times to get $\ket{\vec{0}^{(k)}}$. To get a Z-encoded QR code we instead start with three copies of $\ket{+}$ and execute a $\mathcal{M}^{(Z)}$ gate.  
\item Preparation of $\ket{0^{(k)}_L}$ and $\ket{+^{(k)}_L}$.
By starting with a $3\times3$ array of $\ket{+}$ of physical qubits, and applying a $\mathcal{M}^{(X)}$ in every column we can prepare a $\ket{+_L}$. Similarly $\ket{0_L}$ is obtained by starting with a $3\times3$ array  
of $\ket{0}$ and executing a $\mathcal{M}^{(Z)}$ in every row. This process can be concatenated to fault-tolerantly create encoded $\ket{0^{(k)}_L}$ or $\ket{+^{(k)}_L}$ states at any level of concatenation of concatenation $k$.
\end{itemize}
So this allows us to continuously execute massively parallel unitary fault-tolerant quantum error correction on all logical planes which provides a fault-tolerant memory.

\subsection{Error considerations}
\label{errormodel}
So far we have described a way of executing simultaneous UQEC in every plane of the array. However to truly achieve fault-tolerance we must have a consistent error model  which takes into account that the vertical pulses may introduce correlated errors. We are considering every vertical pulse as a single error location: each possibly faulty vertical addressing pulse admits any error, correlated or uncorrelated, in any of the qubits it addresses. The $semi-global$ character of our design demands also that: errors induced by an $s$-qubit semi-global operation addressing columns $\{(x_1,y_1),..., (x_s,y_s)\}$ are independent of errors induced by an $s'$-qubit semi-global operation addressing columns $\{(x'_1,y'_1),..., (x'_s,y'_s)\}$. Thus in practice we have an adversarial (local) stochastic error model in every plane (in two dimensions).

When we extend our dimensionality, then every $s$-qubit gate addresses $s$ strings of $N_z$ qubits. Let us now assume that our semi-global columnar controls are not applied homogeneously along the individual columns and model an inhomogeneous version of gate a $U$, acting on some column, as $\tilde{U} =  U \prod_z  e^{i \theta_z H_z}$, where $e^{i \theta_z H_z}$ accounts for a potential inhomogeneity of the pulse generated by $H_z$, and  $H_z$ indicates that each location $z$ of the column may undergo a different undesired rotation. Clearly a gate can fail in other ways due to the coupling of the qubits with their environment, but in general we can parametrize such error by some parameter $\theta_z$. So we can model an error in a semi-global $s$-qubit gate as the error implied by the stochastic error model on a $N_z \times s$ qubit columnar gate. For example, a single qubit semi-global gate can be modeled as a $N_z$-qubit physical gate where we assume any type error is possible: single qubit errors, unwanted correlations arising among the $N_z$ qubits, etc. Now the condition for fault-tolerance, under the stochastic error model, is translated into a constraint for every $z$, i.e. that $\theta_z$ is below some threshold $\forall \,\, z$, i.e. in every plane. Note that these sorts of errors are potentially correlated. However we note that these correlations are spatially bounded, i.e. in the case of a $s$-qubit semiglobal gate, are always confined to $s$-qubits in every logical plane, which makes our model effectively a 2D local stochastic adversarial error model. We next extend these error considerations to the vertical $CZ^{(o-e)}$ and $CZ^{(e-o)}$ operations we use to execute universal quantum computation, we assume that an error in $ \mathcal{T}_{(x,y)}$ is independent of an error in $\mathcal{T}_{(x',y')}$. 

As should be expected, our concept to encode separate logical qubits in separate x-y planes to protect them from correlated errors will eventually fail if the error rates are high enough. When enough errors accumulate such that one qubit at the highest level of concatenation is compromised, then all logical qubits at this level of concatenation in the register are compromised since they will eventually couple to it via the spatial reflection protocol, and the computation fails. This may seem a deal breaker but this is no different to what happens in individually addressed fault-tolerance. To show this let's consider a quantum computer running a specific circuit: if at some point enough errors accumulate such that {\it one} of the logical qubits at the highest concatenation level is compromised, then whatever circuit you operate with this logical qubit will be faulty because of error propagation. Thus loosing one or all logical qubits at the highest level of concatenation to accumulated errors is equivalent. 

Additionally, since we plan to use the semi-global control scheme to perform a computation, every logical gate (at the highest concatenation level) is a sequence composed of at least $4(N_z+1)$ $\mathcal{T}$ pulses plus, non-Clifford gates executed at the ends (see next section for details). Thus the computational size of any given circuit we want to simulate is increased by a factor $ \sim 4 (N_z)$, i.e. its' complexity is increased by one order in $N_z$, as compared to the circuit with full addressability. This, as we will see below, has consequences, not in the threshold value, but on the degree of concatenation required to achieve some fixed accuracy in the outcome of a given quantum circuit to be simulated. 

\section{Holographic fault-tolerant universal QC}
\label{HFTUQC}
The previous section described how to execute fault-tolerant quantum error correction using semiglobal control and a compatible error model. Using the scheme in \cite{FitzsimonsTwamley}, we can achieve universal QC, if we can execute  non-Clifford gates at the $z$ end-planes of our 3D array ($z=1$ and $z=N_z$). So, under the error considerations and using the unitary quantum error correction tools described above, if we can execute encoded fault-tolerant Clifford operations at the boundaries we will achieve holographic semi-global fault-tolerant universal quantum computing via $\mathcal{T}$ pulse sequences. Note that after each of the stages in the $\mathcal{T}$ steps (see \eqref{ftT}) we are able to execute unitary quantum error correction. We show now how to achieve the execution of the non-Clifford gates  at the end z-planes. Recall that we allowed the execution of any operation at the boundaries of the array, including measurement, thus we can execute the following routines which will yield the sought fault-tolerant universal quantum computation:
\begin{itemize}

\item[i.] 
{\it $Z^{1/4}$ (non-Clifford) gate aided by encoded measurements.-}
To execute a $Z^{1/4}$ gate we use the circuit in Fig.~\eqref{pi4}. This circuit assumes the ability to prepare a Magic state ancilla, $\ket{H_L}= (\ket{0_L} + e^{i \pi/4}\ket{1_L})/\sqrt{2}$, at the highest level of concatenation with an error rate of the same order of the gates as such level of concatenation. To do so we use a two step routine. First we encode a physical $\ket{H}$ state and use the encoder circuit (described below in (iii)) to get $\ket{H_L}$. This encoded ancilla will typically have an error rate higher than the physical error rate and thus is not yet useful. However it can be shown~\cite{noisymeas} that if the effective error rate of such preparation is below $p_{H-ancs} = 14.6\%$ then they can be used as a resource for magic state distillation~\cite{Bravyi:05}~(MSD). The MSD protocol, composed exclusively of Clifford operations, has two requirements: (i) Clifford operations are perfect, an approximation justified becasue we are executing encoded gates at the highest level of concatenation $L$ and thus Clifford gates can be made arbitraly noiseless provided we choose $L$ large enough, and (ii) a source of noisy copies of $\ket{H_L}$ ancillas with an error rate below $sin^2 (\pi/8) \sim 14.6\%$.

\begin{figure}[h]
\subfigure[]{\includegraphics[width=60mm]{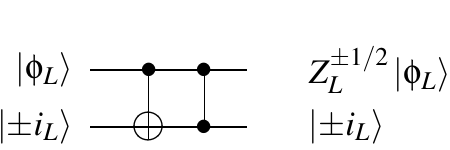}
\label{zhalf}
}
\subfigure[]{
	\centering
		\includegraphics[width=60mm]{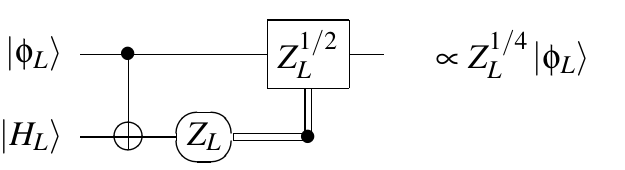}
\label{pi4}	}
	\caption{These circuits need only be implemented at the highest level of concatenation, and thus all operations depicted are encoded operations.  (a) Circuit used to execute an encoded $Z^{1/2}$ gate on an arbitrary input $\ket{\phi_L}$~\cite{Dennis:02} . (b) Circuit implementing the non-Clifford operation $Z^{1/4}$ given the encoded resource magic state $\ket{H_L}$.}   
\end{figure}

\item[ii.] {\it $Z^{1/2}$ (Clifford) gate aided by encoded measurements.-}
Clifford operations can be generated by the gate set $\{ CNOT, H, Z^{1/2}\} $. $CNOT$ and $H$ are transversal, and the $Z^{1/2}$ gate can be implemented using  the circuit in Fig.~\eqref{zhalf}, provided one can  
prepare a logical ancilla in $\ket{\pm i_L} = (\ket{0_L} \pm i \ket{1_L})/\sqrt{2}$. Since the $Z^{1/2}$ gate is not part of the EC routines, it is only needed at the highest level of concatenation. There are two ways of implementing it: (a) since it is the only complex gate, it can be shown that by always using the same logical ancilla prepared in $\ket{0_L} =1/\sqrt{2}(\ket{+i_L} + \ket{-i_L})$ to activate the circuit in Fig.~\eqref{zhalf}, then the entire quantum computation splits into two non interfering paths (evolution by $U_{comp}$ and $U_{comp}^{\ast}$) and the measurements of real, Hermitian operators at the end have the same expectation values as for evolution by $U_{comp}$ alone~\cite{Dennis:02}; or, alternatively, (b) one can use the distillation circuit in \cite{Aliferis} at the highest level provided one can prepare noisy $\ket{+i_L}$ with an error rate below $p_{(i-anc)} = 1/2$.

\item[iii.]{\it Encoding of arbitrary ancillas.-}
Both (i) and (ii) use the preparation of particular encoded states, but they do not require them to have an extremely low error rate. So we will show now a routine to encode any desired state to any level of concatenation, we showed in Ref.\cite{noisymeas}, and summarize in section \eqref{secthresh}, that the error rate of the resulting encoded state $p_{anc}$ is below the required values $p_{(H-anc)}$ and  $p_{(i-anc)}$ and thus can be used as a resource in the routines described above((i.) and (ii)). Thus to encode an arbitrary state we use the following algorithm: (i) we start with the level-0 state $\ket{\phi}$ we want to encode and 8 $\ket{0}$ states, then (ii) we use CNOT gates, including waiting times such that never in one step does one qubit interact with more than one qubit, to create the state  $\ket{\vec \phi}_{3\times3}= a\ket{\vec 0}_{3\times3} + b \ket{\vec 1}_{3\times3}$. Finally (iii) we  execute a $\mathcal{M}^{(Z)}$ gate in every row, to create the state $\ket{\phi_L}= a\ket{0_L} + b \ket{1_L}$. We can recursively use the same algorithm to create the state at any level of concatenation $k$. 
\end{itemize}
Thus combining  the above elements (i)--(iii) we can execute any Clifford and non-Clifford encoded operation in the boundaries. This, in addition to the encoded $\mathcal{T}$, yields a method to execute fault-tolerant $semiglobal$ universal quantum computation. 
\begin{itemize}
\item [iv.] 
The last element of our scheme is the readout stage: to extract the result of a computation. The challenge is to perform the readout without violating the addressability constraints. To do so we use the above described semiglobal FTUQC to execute SWAP gates at the highest level of concatenation to move any $(x-y)$ logical plane  to one of the end planes $(z=1, N_z)$ where measurements are allowed. We can repeat this process until all the desired logical qubits are readout.  
\end{itemize}

\subsection{Restriction to two dimensional arrays}
\label{sec2D}
The tools presented above were based on a three dimensional array of qubits. Despite being natural for many physical systems, it is not even possible for many others. For these latter systems a 2D architecture would be more desirable and we now discuss such a structure. We can start by reducing the dimensionality of every plane to a 1D linear array such that the 3D array is reduced to a 2D planar array. Although this is in principle possible the question is whether fault-tolerance is respected. This depends on the interactions at our disposal in the dimensionally reduced architecture. After the reduction of dimensionality, we now require that we can execute fault-tolerant EC in every line: this is not possible in general if we restrict ourselves to nearest-neighbor interactions only~\cite{GottLayout}, although see~\cite{nnlineFT} for a special case, but is always possible if we also allow next-to-nearest neighbor interactions. We essentially need to show that a SWAP gate between two qubits containing data/information can be executed in a fault-tolerant way, i.e. such that one gate error does not generate more than one error in the data qubits. 

\begin{figure}
	\centering
		\includegraphics[width=0.60\textwidth]{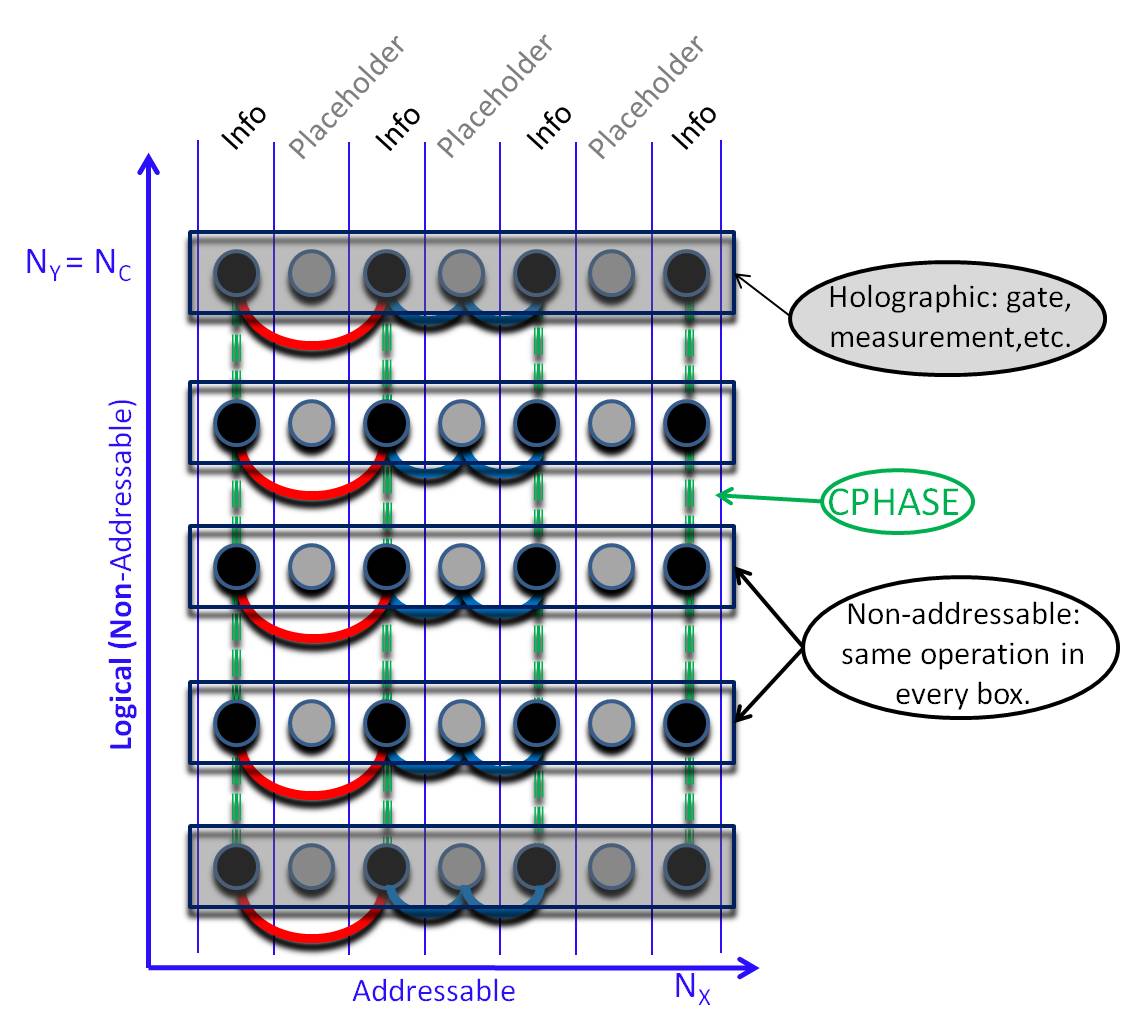}
	\caption{Schematics of the two dimensional architecture. All nearest neighbor interactions are required but only next to nearest-neighbor interactions between odd qubits, i.e. the equivalent to nearest neighbor between info-qubits, are required. This ``cross-free'' structure of the interactions would be beneficial for instance in solid-state implementations since wires would not need to cross. Each line can also be viewed as a one dimensional of 2-type of qubits (ABAB addressability), where only AA, BB and AB nearest neighbor interactions are required. {Recall that in the logical direction we require ABAB addressability to ensure fault-tolerance.}}
	\label{2d}
\end{figure}

To achieve a fault-tolerant swap gate we introduce place-holder qubits, that is qubits that we require specifically to physically move information around but which hold no computationally valuable information. To differentiate them from place-holder qubits, we shall label the qubits involved in the computation, i.e. data plus ancilla qubits, information holding qubits, i.e. info-qubits.  Consider a one dimensional line of $info$-qubits, for example encoding a logical qubit, denoted by $\rho_i$ with interspersed placeholder ancilla qubits, denoted by $\eta_j$, in between every nearest neighbor pair of $info$ qubits (e.g. any horizontal line of Fig.\eqref{2d}). Using nearest neighbor and next-to-nearest neighbor interactions we can now SWAP two $info$ qubits containing relevant information ($\rho$ and $\bar \rho$ respectively), in such a way that a single failure in a SWAP gate does not generate two errors in the $info$ qubits, through the following routine
\begin{center}
$
\begin{array}{c|c|c}
\textrm{Step} & \textrm{Gate} & \textrm{Output state} \\ 
\hline
0 & -& \rho_1 \otimes \eta_{2} \otimes \bar \rho_3 \otimes \eta_{4} \\
1 & SWAP_{1,2}& \eta_1 \otimes \rho_{2} \otimes \bar \rho_3 \otimes \rho_{4} \\
2 & SWAP_{1,3}& \bar\rho_1 \otimes \rho_{2} \otimes \eta_3 \otimes \eta_{4} \\
3 & SWAP_{2,3}& \bar \rho_1 \otimes \eta_{2} \otimes \rho_3 \otimes \eta_{4} 
\end{array}
$
\end{center}
where the subscripts in $\rho$ and $\eta$ denote the physical locations in the sample four qubit chain. This routine executes the fault-tolerant SWAP gate between site 1 and site 3, since qubits containing the relevant information ($\rho$ and $\bar \rho$) never interact directly and a SWAP gate does not propagate errors. Since we can reproduce this structure in a longer chain and achieve SWAP gates between any two qubits in a chain it is possible to execute fault-tolerant semi-global quantum computation on a 2D array with a reduced number of controls. We must have in mind that, although the threshold value will be reduced because of the need to swap qubits around to execute the desired gates, a threshold value does exist~\cite{GottLayout}. It will still remain true that preparation and measurement can be very noisy, as the argument behind such result is that gate error rates are sufficiently below the threshold value. Moreover, the advantage in terms of number of controls is also maintained in this reduced dimensional design.

\section{Holographic semiglobal fault-tolerant threshold}
\label{secthresh}
Let us now discuss the required error rates to achieve fault-tolerant holographic semi-global universal quantum computation. Because we are using the tools developed in Ref.~\cite{noisymeas} we have essentially the same threshold analysis.  
\begin{figure}[htbp]
	\centering
		\includegraphics{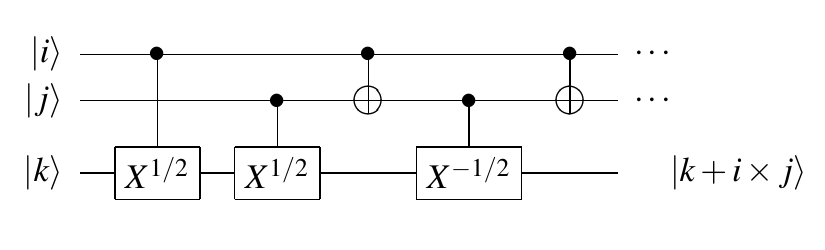}
	\caption{Decomposition of a physical TOFFOLI gate into two qubit interactions when one desired to constraint the problem at most two-qubit interaction only. Note that it requires only three time-steps as the first two $CX^{1/2}$s can be executed simultaneously. The last CNOT gate is not necessary as we will typically discard the controls of such TOFFOLI and thus they do not count towards our threshold estimation. These gates are not encoded gates but always physical gates.}
	\label{extTOFF}
\end{figure}
There we assumed that at the physical level we had at our disposal three qubit gates, in form of TOFFOLI gates. We showed that armed with three-body interactions (in every plane) one can achieve fault-tolerant universal quantum computing if gates and preparation have error rates below $p_{(p,g)thresh} = 3.76\times 10^{-5}$. Measurements are only required at the boundaries and only at the highest level of concatenation, and thus fault-tolerance is possible when measurement error rates are below $p_{(m)thresh} =1/3$. If we implement a gate library at the physical level which does not include the TOFFOLI, then we decompose the TOFFOLI into one and two qubit gates, as in Fig.\eqref{extTOFF}. In this case the threshold value for gates and preparation becomes $p_{(g,p)threshold} = 2.68 \times 10^{-5}$ again with measurements as noisy as $ p_{(m)} = 1/3$. 

Furthermore, we showed in ~\cite{noisymeas} that provided that the gate error rate ($p_{(g)}$) was sufficiently below $p_{(g)threshold}$ then one could relax the demands on measurement and preparation error rates, ($p_{(m)}$) and ($p_{(p)}$) respectively. Following that argument, and assuming our gate library contains the TOFFOLI, we find that if $p_{(g)} = 1.3 \times 10^{-6}$, $p_{(p)} = 1 \% $ and $p_{(m)} = 33\%$ yields, with e.g. $k=6$ degrees of concatenation, and effective error rate at concatenation level $k=6$, $p^{(6)} \sim 10^{-13}$ and and effective error rate for the output of an encoded state, $p^{(6)}_{anc} = 8.3 \times 10^{-3}$.  Note that $p^{(6)}_{anc}$, denoting the ouput error rate of the $\ket{H_L}$ encoder circuit, is safely below the $1.46 \times 10^{-1}$ needed to execute magic state distillation and achieve fault-tolerant universal quantum computation. 

This implies that the massively parallel mapping required to refresh the ancillas in our design can have error rates $\tilde p_{(p)}$ as high as $1\%$, or even $33\%$ at the expense of demanding even lower values from $p_{(g)}$ ($
\sim 10^{-6}$). To accommodate more physically realistic interactions and assume only nearest neighbor interactions in the $xy$ planes then one can expect a decrease of the threshold. However this is a characteristic shared by any addressable or non-addressable fault-tolerant scheme, and previous work has shown that by restricting a long-range addressable fault tolerant scheme to be nearest-neighbor can decrease the threshold by less than an order of magnitude~\cite{nnSteaneandBS}. Further, we observe that the dynamical decoupling (DD) protection of gates still applies, and such DD protection can be made compatible with our reduced addressability~\cite{LidarDD}. It follows that the extra demand placed on gate error rates($p_{(g)} \sim 10^{-5} - 10^{-6}$) can in principle be greatly alleviated by open quantum system control techniques~\cite{NewLidar}.

\section{General strategy} 
\label{controlsec}
The particular strategy to achieve semi-globally addressed fault-tolerant QC we presented above may not be unique, so here we want to summarize what are some of the essential requirements of our design. Our scheme relies on two properties: (i) every plane can execute error correction in a fault-tolerant manner, and (ii) the $z$ direction is in charge of the computational aspect of our array via some interaction capable of coupling different planes. We have chosen the $\mathcal{T}$ pulse, but in principle other scenarios inspired by other global control strategies could be implemented. 

Lets say we have a global control scheme in one dimension with some (not necessarily nearest-neighbor) interaction $\mathcal{T}$. We consider a 3D array as a collection of 2D collection of 1D arrays, such that at every 1D array we can execute $\mathcal{T}$ independently. Every horizontal plane of our 3D array will constitute a logical qubit, encoded in some QEC code with a set of stabilizers $\{ S^{(z)}_i \}$, and the necessary ancillas for its measurement-free error correction. This means that our computer will be initially stabilized by $\bigotimes_{1}^{N_z} \{ S^{(z)}_i \}$. This fixes a relationship between  $\mathcal{T}$ and $\{ S^{(z)}_i \}$. We must guarantee that (i) at all times all logical qubits in the computer must be stabilized by some (possibly changing) code $\{ S^{(z)}_i \}$, i.e. $\mathcal{T}$ never leaves the information unprotected, and (ii) $\mathcal{T}$ can propagate an error in some plane to only one qubit in any number of planes (nearest neighbor interactions fix the propagation of errors to nearest neighbor planes). The above requirements will be enough to achieve parallel FT UQEC in all qubits and transport of information. Extra requirements for fault-tolerant universal computation depend on the type of global control scheme one considers.

\subsection{Scaling of the number of controls}

To see that the semi-global architecture saves on resources we now investigate the spatial addressing efficiency of our design. We first count the number of controls required to execute a fault-tolerant semi-global quantum computer architecture and compare it with a fully addressable quantum computer architecture simulating the same quantum circuit to the same overall accuracy. As a first step, we assume the same measurement free EC routines but with full addressing, since we really want to know if we gain something using the 3D layout instead of just using a 2D one, and compare the number of controls required ($N_{[uAdd]}$ and $N_{[sg]}$ respectively) . We then proceed to compare the 3D layout to a fully addressable, measurement capable architecture using the same QEC code (which will be labeled by [mAdd]).

Defining the parameters: $N_C$ logical/computational qubits encoded using a QEC code which encodes each logical qubit into $N_{EC}$ physical qubits with $k$ degrees of concatenation, and EC gadget using $N_A$ encoded ancilla qubits and $N_B$ classically encoded ancilla qubits. A naive count shows that the number of controls in the fully addressable measurement-free (labeled with [uAdd]) and semiglobal measurement free (labeled with [sg]) architectures are given by \footnote{Noting that a level-$k$ encoding is composed of $(N_{EC} + N_A)$ level-$(k-1)$ BS encoded qubits and  $N_B$ level-$(k-1)$ QR encoded qubits.}
\beqy 
N_{[uAdd]} &=&  N_C \times [(N_{EC}+ N_A)\times 9^{{k_{[uAdd]}}-1} + N_B \times 3^{{k_{[uAdd]}}-1}]     \\
\label{controls}
N_{[sg]} &=&  (N_{EC}+N_A)\times 9^{{k_{[sg]}}-1}+ N_B \times 3^{{k_{[sg]}}-1}
\eneqy
Using the EC protocols described before, $N_{EC} = 9, N_A = 18$ and $N_B = 6$.  From \eqref{controls} it would seem that $N_{[sg]}$ does not depend on the number of computational qubits, and that $N_{[uAdd]} / N_{[sg]} = N_C$,  however that is not the case as $k_{[sg]}$ is a function of $N_C$, albeit with a weaker dependence. To see this we consider the result of the threshold theorem~\cite{threshold}: given a circuit which we wish to simulate to an accuracy $\epsilon$, whose size is a polynomial in the number of computational qubits, $f(N_C)$, then 
\beqy
p^{(k)} \times f(N_C) &\leq& \epsilon  \\
(A p^{(0)})^{2^{k}}/ A  &\leq& \epsilon /f(N_C),
\eneqy
where $p^{(k)} = A (p^{(k-1)})^2$ is the error probability of an operation at level $k$ of concatenation in terms of level-(k-1) error rates and $A$ counts the number of pairs of possible level-$(k-1)$ errors (in the largest exRec defined in \cite{Aliferis} as a gate with appended and preprended $EC(k)$ routines)
\beq
k_{[\cdot]}  = \log_2 \left[ \frac{\log \left(\frac{\epsilon/p_{thresh}}{f_{[\cdot]}(N_C)} \right)}{\log \left( p^{(0)}/p_{thresh}\right)}  \right], 
\eneq
where we have emphasized the dependence of $f(N_c)$ on the design, i.e. for $[\cdot] = \{ [sg], [uAdd], [mAdd] \}$. This equation gives real values whenever concatenation is not harmful. Assuming that the polynomial scaling of the circuit size with the number of computational qubits is with power $t$, i.e. $f_{[uAdd]}(N_C) = \beta N_C^t$, we have
\beqy
\delta k \equiv k_{[sg]} - k_{[uAdd]} &=& \log_2 \left[ 1 + \frac{\log(4 N_C)}{\log( f_{[uAdd]}(N_C) p_{thresh}/ \epsilon)}\right]\\
\nonumber  &=&\log_2 \left[ 1 + \frac{\log(N_C) + \log{4}}{\log N_C + \log (\beta N_C^{t-1} p_{thresh}/ \epsilon))}\right]
\eneqy
In general, because of the discreteness of $k$ we will have an effective discrete difference between degrees of concatenation $\Delta k \geq 0 $ instead of a continuous value for $\delta k$. This discrete difference oscillates as the number of computational qubits increases. In any case we will have that 
\beqy 
N_{[uAdd]}/N_{[sg]} &=& N_C \frac{(N_{EC}+ N_A)\times 9^{k_{[uAdd]}} + N_B \times 3^{k_{[uAdd]}}}{(N_{EC}+ N_A)\times 9^{k_{[sg]}} + N_B \times 3^{k_{[sg]}}}\\
&=& N_C \frac{1}{3^{\Delta k}} \left( \frac{2 + 3^{2+k_{[uAdd]}}}{ 2 + 3^{2+k_{[sg]}}}  \right),
\eneqy 
where $\Delta k = \lceil k_{[sg]} \rceil - \lceil k_{[uAdd]} \rceil$. So in general the semiglobal architecture will have a gain $\propto N_C$ in the number of controls, i.e. $N_{[uAdd]}/N_{[sg]}>1$. More generally, we determine that there is a significant advantage, i.e. $\Delta k \leq 1 \Rightarrow N_{[uAdd]}/N_{[sg]} =N_C$, whenever $\log{4} \leq \log{(\beta N_C^{t-1} p_{thresh}/ \epsilon)}$. 

As an example, let us consider the Shor factorization algorithm. To factor a $N=768$ bit integer would require $N_C = 2 N  + 4 = 1540$ qubits, and a circuit size of $8 N^4 = 2.8 \times 10^{12}$ logical gates, if we want a 97\% overall success rate, using $ p^{(0)} = 10^{-6} < p_{thresh}$ we get that $\Delta k = 1$ and thus we gain factor of $N_{[uAdd]}/N_{[sg]} \sim 40.6 \times 10^6 / 19845 =  768 $ reduction in the number of controls controls. A N=2048 string would yield $\Delta k = 0$ and thus a gain of $N_{[uAdd]}/N_{[sg]} \sim 4.59 \times 10^6 /19845 = 2048$, while a 4096 string yields $\Delta k = 1$ and $N_{[uAdd]}/N_{[sg]} \sim 81.3\times 10^6 / 19845 \sim 4096$.  

We can go a step further and compare a fully addressable model with error correction gadgets that admit measurements with our  semi-global unitary QEC model. We will consider the gadgets and threshold values for our same QEC code obtained in Ref.\cite{Aliferis} with a corresponding set of parameters $N'_A=18, p'_{threshold}= 1.2 \times 10^{-4}$. Because such EC gadgets would typically use a different number of ancillas $N'_{A} < N_A$ and have different threshold value $p'_{threshold}$, we get \beq 
N'_{[mAdd]}/N_{[sg]} = N_C \frac{(N_{EC}+ N'_A)\times 9^{k'-1}}{(N_{EC}+ N_A)\times 9^{k_{[sg]}-1} + N_B \times 3^{k_{[sg]}-1}}
\eneq
with 
\beqy
\delta k' &=& k_{[sg]} - k'\\
\nonumber &=& -\log_2{\left[ \frac{\log(\epsilon ( p'_{thresh} f'_{[mAdd]}))}{\log(\epsilon/( p_{thresh} f_{[mAdd]})) }  \times \frac{\log (p^{(0)}/p_{thresh})}{\log (p^{(0)}/p'_{thresh})}  \right]}. 
\eneqy
The difference with the previous case is that now $\Delta k'$ is more likely to be 1 and not 0, as follows from the observation that typically $k'<k_[uAdd]$. Assuming we have a physical error rate good enough for both threshold values, $p_{thresh}$ and $p'_{thresh}$, e.g. $p^{(0)} = 1 \times 10^{-6}$, we get for Shor's protocol on the strings used in the previous example, $ \Delta k'= 1$ and $N'_{[mAdd]}/N_{[sg]} |_{768} \sim 1.68 \times 10^6 / 19845 \sim  84$, for 2048 $\Delta k'= 1$ and thus $N'_{[mAdd]}/N_{[sg]} |_{2048}~ 4.48\times 10^6 / 19845  \sim 225$, and for 4096 $\Delta k'= 1$ again and we get $N'_{[mAdd]}/N_{[sg]} |_{4096} \sim 8.96 \times 10^6/ 19845 \sim 451$ which implies a slightly smaller gain than in the previous case, but still significant $\sim N_C ( \frac{1}{ 9^{\Delta k'} + 3^{\Delta k'- k'+1}\times 2/3})$. Even compared to gadgets where measurements are allowed the semiglobal architecture yields a $\mathcal{O}( N_C)$ gain in terms of the number of controls.   

\section{Experimental considerations}

In terms of an experimental realization, given a 3D array of qubits, the addressability requirements can be translated into :
\begin{itemize}
\item[i.] Massively parallel nearest neighbor CZ gates in the z direction: $\prod_{(x,y)} \tilde T_{(x,y)}$.
\item[ii.] {\it Single-qubit} (horizontal) gates: $U_{(x,y)} = \prod_{z} U_{(x,y,z)}$ for any $(x,y)$
\item[iii.] {\it Two-qubit} (horizontal) gates: $V_{((x,y),(x',y'))} = \prod_{z} V_{(x,y,z),(x',y',z)}$, for any $(x,y)$ and $(x',y')$ nearest neighbors in a 2D lattice. 
\item[iv.] Massively parallel resetting: $\rho \rightarrow \ket{0}\bra{0}$
\end{itemize}

One candidate technology for quantum computation well suited to holographic control is neutral atoms trapped in a three dimensional optical lattice.  In this architecture one has the advantage of massive parallelism but since the lattice spacing is typically on the order of an optical wavelength it is difficult to accommodate imaging lenses that could resolve individual qubit measurement outcomes in the bulk.  To get some idea of the size of a computation that could be realized in such a system, in Ref.  \cite{Vala} the authors find that $10^6$ ${^{133}}$Cs atoms could be trapped in a $100\times100\times 100$ blue detuned lattice with an achievable single qubit gate error rate of $10^{-5}$ using Raman based gates.  Here the lattice spacing would be $10\mu m$ and it has already been demonstrated \cite{Greiner} that single 1D tubes of ${^{87}}$Rb atoms trapped in a 2D lattice can be addressed with better than $1\mu m$ resolution.  Furthermore, using an architecture such as the 3D retroreflected lattice used in \cite{Anderlini} it is possible to have $ABAB$ type addressability along one or two dimensions enabling addressability of every other plane.   If we make the reasonable assumption of a single qubit reset error rate of $10^{-5}$ and the (extremely optimistic) assumption of the same error rate for two qubit gates, then restricting to nearest neighbour interactions, a computation with $\sim 10^5$ sequential logical gates could be achieved on $100$ logical qubits using holographic control with an overall circuit simulation error of $1/3$.  This would require $3$ levels of concatenation which could be accommodated in each $100\times 100$ plane with room left over for resettable ancilla to shuttle quantum information.  Massively parrallel two qubit CPHASE gates have been realized in optical lattices but with rather low fidelity \cite{Mandel}.  The best reported two qubit gate using parallel exchange blockade mechanism between neighboring trapped atoms in an optical lattice realised an error for the $\sqrt{{\rm SWAP}}$ gate of $0.31$ \cite{Anderlini} though theory predicts error rates as low as $1\%$ could be achieved \cite{Hayes}.  There are several other proposals for high fidelity entangling gates in optical lattices, e.g. using fast Rydberg gates \cite{Zoller}, but to achieve an error below our threshold would likely require another approach such as using dynamical decoupling pulses to boost the effective two qubit gate fidelity \cite{NewLidar}. \\

We note that the energy required by a control pulse increases with the number of x-y planes it must control and this may pose a limit to the size of the computation the architecture may implement. However if the physical model permits, one could place several several logical qubits per plane, i.e. one plane = several tiles of logical qubits, such that a pulse with limited addressing capacity can still be used to build an, in principle, arbitrarily large computer with the tools described in this paper. Universality follows form the fact that we can achieve SWAP gates between tiles within different planes and thus execute any two logical qubit gate.

\section{Conclusions}

We have shown that in a N-dimensional ($N=2,3$) qubit array fault-tolerance is achievable when only $(N-1)$-dimensional addressability and fixed short-range interactions are available. The scheme has implications on the design of scalable quantum computers and shows an advantage in the number of controls required to manipulate the array of qubits. More specifically the number of controls required depends only weakly on the number of computational qubits, as opposed to fully addressable designs where they grow linearly, or equivalently we have a gain factor $\mathcal{O} (N_C)$ in the number of controls. The design is suitable for 3D optical lattices, and for 2D arrays with only nearest and next-to-nearest neighbor couplings.

\acknowledgements
GKB received support from the EC project AQUTE and from the Australian Research Council. JT acknowledges support from Q-ESSENCE. GAPS acknowledges support from Macquarie University and CQCT.


\begin{thebibliography}{99}
\bibitem{Shor} P. W. Shor, SIAM J. Comput. {\bf  26} (5), 1484 (1997).
\bibitem{DiVincenzo} D. P. DiVincenzo, Fortschritte der Physik {\bf 48}, 771 (2000).
\bibitem{QEC} P. W. Shor, Phys. Rev. A {\bf 52}, R2493 (1995).
\bibitem{threshold} D. Aharonov and M. Ben-Or, SIAM J. Comput. {\bf 38}, 1207 (2008); arXiv:quant-ph/9906129.
\bibitem{IonTap} J. I. Cirac and P. Zoller, Phys. Rev. Lett. {\bf 74}, 4091 (1995), D. Kielpinski, C. Monroe and D. J. Wineland, Nature {\bf 417}, 709 (2002)
\bibitem{OLatt} G. K. Brennen, C. M. Caves, P. S. Jessen, and I. H. Deutsch, Phys. Rev. Lett. {\bf 82}, 1060 (1999) 
\bibitem{noisymeas} G. A. Paz-silva, G. K. Brennen and J. Twamley, arXiv:1002.1536. To appear in Phys. Rev. Lett.
\bibitem{NatureHardMeasurements} T. D. Ladd { \it et al.}, Nature {\bf 464}, 45 (2010); H. J. Briegel {\it et al.}, Nature Physics , 19 - 26 (2009). 
\bibitem{FitzsimonsTwamley} J. Fitzsimons and J. Twamley, Phys. Rev. Lett. {\bf 97}, 090502 (2006)
\bibitem{Raussendorf} R. Raussendorf, Phys. Rev. A { \bf 72}, 052301 (2005).
\bibitem{GBanacloche} J. P. Clemens, S. Siddiqui, and J. Gea-Banacloche, Phys. Rev. A {\bf 69}, 062313 (2004)
\bibitem{Dennis:02} E. Dennis {\it et al.}, J. Math. Phys. {\bf 43}, 4452 (2002).
\bibitem{Aliferis} P. Aliferis, Ph.D. thesis, Caltech, 2007; arXiv:0703230 and references therein.
\bibitem{FTGCthresh} J. Fitzsimons and J. Twamley, Electronic Notes in Theoretical Computer Science { \bf 258 } (2), 35 (2009).
\bibitem{CSS} A. R. Calderbank and P. W. Shor, Phys. Rev. A {\bf 54} (1996), 1098; A. M. Steane, Proc. Roy. Soc. Lond. A {\bf 452}, 2551 (1996). 
\bibitem{Bravyi:05} S. Bravyi and A. Kitaev, Phys. Rev. A {\bf 71}, 022316 (2005); B. Reichardt, Quant. Inf. Comp. {\bf 9}, 1030  
(2009).
\bibitem{GottLayout} D. Gottesman, J.Mod.Opt. {\bf 47}  333 (2000).
\bibitem{nnlineFT} A. M. Stephens and Z. W. E. Evans, Phys. Rev. A {\bf 80}, 022313 (2009)
\bibitem{nnSteaneandBS} K. M. Svore et al., Quantum Inf. Comput. {\bf 7}, 297 (2007); F. M. Spedalieri and V. P. Roychowdhury, Quantum Inf. Comput. {\bf 9}, 0666 (2009).
\bibitem{LidarDD} H. K. Ng et al., arXiv:0911.3202; J.R. West et al., arXiv:0911.2398v1.
\bibitem{NewLidar} K. Khodjasteh, D. A. Lidar, and L. Viola, Phys. Rev. Lett. {\bf 104}, 090501 (2010); K. Khodjasteh and L.Viola, Phys. Rev. Lett. {\bf 102}, 080501 (2009); Phys. Rev. A {\bf 80}, 032314 (2009).
\bibitem{Vala} T.R. Beals, J. Vala, and K.B. Whaley, Phys. Rev. A {\bf 77}, 052309 (2008).
\bibitem{Greiner} W.S. Bakr, J.I. Gillen, A. Peng, S. F\"olling, and M. Greiner, Nature {\bf 462} 74 (2009); W.S. Bakr, A. Peng, M. E. Tai, R. Ma, J. Simon, J.I. Gillen, S. F\"olling, L. Pollet, and M. Greiner, arXiv:1006.0754.
\bibitem{Mandel} O. Mandel, M. Greiner, A. Widera, T. Rom, T.W. H\"ansch, and I. Bloch, Nature {\bf 425} 937 (2003).
\bibitem{Anderlini} M. Anderlini, P.J. Lee, B.L. Brown, J. Sebby-Strabley, W.D. Phillips, and J.V. Porto, Nature {\bf 448} 452 (2007).
\bibitem{Hayes}  D. Hayes, P.S. Julienne, I.H. Deutsch, Phys. Rev. Lett. {\bf 98}, 070501 (2007).
\bibitem{Zoller} D. Jaksch, J.I. Cirac, P. Zoller, S.L. Rolston, R. C\^{o}te, and M.D. Lukin, Phys. Rev. Lett. 
{\bf 85}, 2208 (2000).
\end{thebibliography}
\end{document}